\documentclass[aps,twocolumn,amsmath,amssymb,nofootinbib]{revtex4}

\usepackage{graphicx}% Include figure files
\usepackage{dcolumn}% Align table columns on decimal point
\usepackage{bm}% bold math
\usepackage{hyperref}% add hypertext capabilities
\usepackage[mathlines]{lineno}% Enable numbering of text and display math
\usepackage{braket}
\usepackage{amssymb}
\usepackage{multirow}
\begin{document}
\makeatletter
\newcommand{\rmnum}[1]{\romannumeral #1}
\newcommand{\Rmnum}[1]{\expandafter\@slowromancap\romannumeral #1@}
\newcommand{\bs}{\boldsymbol}
\makeatother

\title{Possible Assignment of Excited Light $^3S_1$ Vector Mesons}

\author{Jie-Cheng Feng,$^{1}$ Xian-Wei Kang,$^{2,3}$ Qi-Fang L\"u,$^{4,5,6}$ Feng-Shou Zhang$^{2,3}$}

\affiliation{
$^{1}$Department of Physics, Beijing Normal University, Beijing 100875, China\\
$^{2}$Key Laboratory of Beam Technology of the Ministry of Education, College of Nuclear Science and Technology,
Beijing Normal University, Beijing 100875, China\\
$^{3}$Beijing Radiation Center, Beijing 100875, China\\
$^{4}$Department of Physics, Hunan Normal University, Changsha 410081, China\\
$^{5}$Synergetic Innovation Center for Quantum Effects and Applications (SICQEA), Changsha 410081, China\\
$^{6}$Key Laboratory of Low-Dimensional Quantum Structures and Quantum Control of the Ministry of Education, Changsha 410081, China
}

%\date{\today}% It is always \today, today,
             %  but any date may be explicitly specified

\begin{abstract}
We reanalyze the problems in the assignment of 3$^3S_1$ and 4$^3S_1$ light mesons,
which have not yet been well established with the $q\bar{q}$ quark model. Regge trajectories and
the $^3P_0$ decay model are used respectively to study the mass and width of the observed states and
predict the missing ones. By comparing our calculations with the latest experiments, we suggest that the
inconsistent data of $\rho(2150)$ may include two similar structures $\rho(4^3S_1)$ and $\omega(4^3S_1)$.
In addition, the problem of the $K^*(2^3S_1)$ assignment, with two observed states $K^*(1410)$ and $K^*(1680)$,
is investigated, with several possible explanations.
\end{abstract}
\maketitle

\section{{\label{sec:level1}Introduction}}
The use of spectroscopy to study light mesons (constituents are light quarks) and the search for missing excited states are
long-standing and meaningful topics in hadronic physics. With the abundance of experimental
information available, various properties of these states have been investigated, and the low-lying spectra
for some sectors have been established. For instance, the pseudoscalar nonet up to four radial excitations has
been investigated by several works \cite{Barnes:1996ff,Barnes:2002mu,Li:2008mza,Liu:2010tr,Yu:2011ta,Wang:2017iai,Xue:2018jvi},
and the assignments of low-lying ones, such as the $1^1S_0$, $2^1S_0$, and $3^1S_0$ states, agree with each other. Moreover,
$\eta^{(\prime)}(6S)$ and $\eta^{(\prime)}(7S)$ have also been investigated recently \cite{Wang:2020due}.

Unlike the case of pseudoscalar nonets, the spectra for the light vector nonet are far from being established.
Light vector mesons include four families: the $\rho$ meson for isospin $I=1$; the kaon for $I=\frac{1}{2}$;
and the $\omega$ meson and $\phi$ meson for $I=0$, which are made from $(u\bar{u}+d\bar{d})/2$ and $s\bar{s}$, respectively.
In addition to the ground states and the $\rho(1450)$, $\omega(1420)$, and $\phi(1680)$ resonances, the current assignment of 3$^3S_1$ and 4$^3S_1$,
even a part of 2$^3S_1$, is confusing, especially for the $\phi$ and $K$ families due to limited data \cite{Zyla:2020zbs}.
One of the main purposes of this paper is to examine the possibility of some related states to be explained as $q\bar{q}$ mesons with
definite quantum numbers. If the structure cannot be understood well by the $q\bar{q}$ model, it might be distinguished as exotica,
including glueball, tetraquark and hybrid mesons, or a complex mixture of exotica and conventional mesons.

$e^+e^-$ collision experiments on BES$\text{\Rmnum{3}}$ and BaBar provide data over 2.0-2.2 GeV with high precision,
which allow us to study the excitations of the $\rho$, $\omega$ and $\phi$ families closely
\cite{Ablikim:2018iyx,BABAR:2019oes,Ablikim:2020coo,Ablikim:2020das,Ablikim:2020wyk}. The degeneracy between the masses of
excited $\rho$ and $\omega$ mesons, with the observed $\phi$(2170), indicates that this mass region is rich in resonance.
Therefore, we also try to offer some clues from our calculation to explain the inconsistency of observations of a $\rho$-like
structure with a mass around 2.1 GeV. Another kind of experiment that is worth mentioning is the $p\bar{p}$ collision of the
Crystal Barrel experiment, which has led to the discovery of several candidate excitations of the $\omega$ meson \cite{Anisovich:2011sva,Bugg:2004rj}.
However, these structures lack confirmation by other collaborations and are listed as ``Further States'' by the Particle Data Group (PDG).
The kaon data are scarcer; the properties of $K^*$ mainly come from LASS in the 1980s and LHCb in 2017, resulting in ambiguity
in the assignment \cite{Aston:1986jb,Aston:1987ir,Aaij:2016iza}.

First, we find the possible candidates by comparing the mass predicted from Regge trajectories with the mass from experiments.
Next, the widths of these states, assumed to be $q\bar{q}$, relative to all OZI-allowed two-body decays are calculated with the $^3P_0$ model.
The result should be enough to compare with the total widths from experiments to derive some qualitative, or even quantitative, conclusions.
If the data do not fit our simple model well, we also provide some possible explanations for this phenomenon, in addition to the exotic state.
Several recent works on vector meson assignment focus mainly on one family \cite{He:2013ttg,Wang:2019jch,Pang:2019ttv,Li:2020xzs}; therefore,
we want to present some new insights from the aspect of the nonet.

The organization of this paper is as follows: After we present a brief summary of the $^3P_0$ decay model
and the parameters we adopt in Section~\ref{sec:level2}, the $\rho$ meson and $\omega$ family are discussed in
Section~\ref{sec:level3}, with numerical values given for their decay widths. The following Section~\ref{sec:level4}
presents a similar exercise for $\phi$ and $K$ mesons, the conclusions regarding which, however, are more speculative.
Finally, we end with a short summary of our major conclusions and suggest some interesting topics for future study in Section~\ref{sec:level5}.

\section{{\label{sec:level2}Model and parameters}}
We use the $^3P_0$ model to calculate the OZI-allowed two-body strong decay widths. This model assumes that each hadron decays in a simple way
that a quark pair is created from vacuum with quantum number $0^{++}\,(^3P_0)$ and combination of this quark pair with the ones from mother hadron
constitute the dauther hadrons. The model was proposed by Micu 50 years ago \cite{Micu:1968mk} and then further developed in the 1970s by the
Orsay group \cite{LeYaouanc:1972vsx,LeYaouanc:1973ldf} and has been extensively applied to hadron strong decay since then, e.g.,
Refs.~\cite{Blundell:1995ev,Barnes:1996ff,Capstick:2000qj,Barnes:2002mu,Close:2005se,Zhang:2006yj,Godfrey:2016nwn}. Admittedly, there may be some inaccuracy for predicting the spectrum with the nonrelativistic quark model in a general sense. However, $^3P_0$ model considered here, as only one branch of the nonrelativistic quark model, is used to calculate the hadronic decay width, which is not directly related to those issues. It is actually surprising that this approximated model explains experimental data with considerable success, as shown in the above references. There are some potential improvements of the $^3P_0$ model that could be useful for the assignment of quantum number in the future. For example, some recent works point out that considering both $^3P_0$ and $^3S_1$ mechanisms together may lead to better angular dependence and fit the data more accurately \cite{Bathas:1993zm,Nagels:2014qqa,El-Bennich:2003phi}. More technical details of the $^3P_0$ model can be found in Refs.~\cite{Blundell:1996as,Ackleh:1996yt,LeYaouanc:1988fx}

The general idea of the model is that the initial quarks are spectators as $q\bar{q}$ is created from a vacuum. Similarly,
one quark and one antiquark combine into a meson without affecting the other quarks. To express the simplification quantitatively,
the transition operator $T$ of decay $A\rightarrow B+C$ in the $^3P_0$ model, defined only for the decay process but avoiding the Hamiltonian,
is written as
\begin{eqnarray}
T=&&-3\gamma\sum_{m}\braket{1m1-m|00}\int d\boldsymbol{p_3}d\boldsymbol{p_4}\delta^3(\boldsymbol{p_3}+\boldsymbol{p_4})\nonumber\\
&&\times\mathcal{Y}^m_1(\frac{\boldsymbol{p_3}-\boldsymbol{p_4}}{2})\chi^{34}_{1-m}\phi^{34}_0\omega^{34}_0b^{\dagger}_{3i}(\boldsymbol{p_3})
d^{\dagger}_{4j}(\boldsymbol{p_4})
\label{eq:one}
\end{eqnarray}
The indices 3 and 4 denote the quark and antiquark produced from a vacuum;
$b^{\dagger}_{3i}(\boldsymbol{p_3})$ and $d^{\dagger}_{4j}(\boldsymbol{p_4})$ are creation operators
with SU(3)-color indices $i, j$; $\phi^{34}_0=\frac{u\bar{u}+d\bar{d}+s\bar{s}}{\sqrt{3}}$ and $\omega^{34}_0=\delta^{ij}/\sqrt{3}$
correspond to flavor and color singlets;
$\chi^{34}_{1-m}$ denotes the spin wave function; and $\mathcal{Y}^m_l(\boldsymbol{k})=|\boldsymbol{k}|^lY^m_l(\theta_k,\phi_k)$
is the solid harmonic polynomial. $\gamma$ is a universal
dimensionless parameter reflecting the strength of the creation of the $q\bar{q}$ pair. Note that in Ref.~\cite{Segovia:2012cd},
a scale-dependent $\gamma$ with the logarithm of
the reduced mass of the quark pair is proposed, and the parameters are determined by fitting to some measured total widths.

We can define the transition matrix in a simple form:
\begin{eqnarray}
\braket{BC|T|A}=\delta(\boldsymbol{P_B}+\boldsymbol{P_C})\mathcal{M}^{M_{J_A}M_{J_B}M_{J_C}}(\boldsymbol{P_B})
\end{eqnarray}
where $\boldsymbol{P}_{B(C)}$ is the three-momentum of meson $B$ or $C$ in the rest frame of $A$; ${J_i}$ is the total angular momentum of particle $i$;
and $\mathcal{M}^{M_{J_A}M_{J_B}M_{J_C}}$ is the decay helicity amplitude. From Eq.~(\ref{eq:one}), the explicit form of the decay amplitude is:
\begin{eqnarray}
&&\mathcal{M}^{M_{J_A}M_{J_B}M_{J_C}}(\boldsymbol{P})\nonumber\\
&&=\gamma\sqrt{8E_AE_BE_C}\sum_{\substack{M_{L_A},M_{S_A}\\M_{L_B},M_{S_B}\\ M_{L_C},M_{S_C},m}}\braket{L_A M_{L_A} S_A M_{S_A}|J_A M_{J_A}}\nonumber\\
&&\times \braket{L_B M_{L_B} S_B M_{S_B}|J_B M_{J_B}}\braket{L_C M_{L_C} S_C M_{S_C}|J_C M_{J_C}}\nonumber\\
&&\times \braket{1m1-m|00}\braket{\chi^{14}_{S_B M_{S_B}}\chi^{32}_{S_C M_{S_C}}|\chi^{12}_{S_A M_{S_A}}\chi^{34}_{1 -m}}\nonumber\\
&&\times [\braket{\phi^{14}_B\phi^{32}_C|\phi^{12}_A\phi^{34}_0}I(\boldsymbol{P},m_1,m_2,m_3)\nonumber\\
&&+(-1)^{1+S_A+S_B+S_C}\braket{\phi^{32}_B\phi^{14}_C|\phi^{12}_A\phi^{34}_0}I(\boldsymbol{-P},m_2,m_1,m_3)]\nonumber\\
\end{eqnarray}
The indices in the spin and flavor wavefunction overlaps serve to identify the quarks. $m_1, m_2, m_3$ are the masses of the quarks, and
we use $m_u=m_d=0.33$ GeV and $m_s=0.55$ GeV. The two terms of the last factor correspond to two possible combinations, and
$I(\boldsymbol{P},m_1,m_2,m_3)$
is the integral of the space wave function overlap:
\begin{eqnarray}\label{eq:I}
&&I(\boldsymbol{P},m_1,m_2,m_3)\nonumber\\&&=\int d^3\bs{p}\,\psi^*_{N_B L_B M_{L_B}}\left(\frac{m_3}{m_1+m_3}\bs P+\bs p\right)\nonumber \\
&&\times\psi^*_{N_C L_C M_{L_C}}
\left(-\frac{m_3}{m_2+m_3}\bs P-\bs p\right) \nonumber\\ &&\times\psi_{N_A L_A M_{L_A}}\left(\bs P+\bs p\right)
\mathcal{Y}_1^m(\bs p),
\end{eqnarray}
where in the center of mass frame $\boldsymbol{P}\equiv\boldsymbol{P_B}=-\boldsymbol{P_C}$.

In this paper, we apply the simple harmonic oscillator (SHO) wave function to describe the $q\bar{q}$ meson in the calculation.
$L^{l+1/2}_{n}$ is the associated Laguerre polynomial.
\begin{eqnarray}
\Psi_{nlm}(\boldsymbol{p})&=&\frac{(-1)^n(-i)^l}{\beta^{3/2}}\sqrt{\frac{2n!}{\Gamma(n+l+\frac{3}{2})}}\nonumber\\
&&\times e^{-(\frac{p^2}{2\beta^2})}L^{l+1/2}_{n}(p^2/\beta^2)\mathcal{Y}^m_l(\boldsymbol{p}/\beta)
\end{eqnarray}

The partial wave amplitude $\mathcal{M}^{LS}(\boldsymbol{P})$ can be derived by the Jacob-Wick formula \cite{Jacob:1959at},
\begin{eqnarray}
&&\mathcal{M}^{LS}(\boldsymbol{P})=\nonumber\\&&\frac{\sqrt{4\pi(2L+1)}}{2J_A+1}\sum_{M_{J_B}M_{J_C}}\braket{L0SM_{J_A}|J_AM_{J_A}}\nonumber\\
&&\times\braket{J_BM_{J_B}J_CM_{J_C}|S M_{J_A}}\mathcal{M}^{M_{J_A}M_{J_B}M_{J_C}}(\boldsymbol{P}\hat z)
\end{eqnarray}
and in fact, $\boldsymbol{P}=\boldsymbol{P}_B$ is chosen as along $z$-axis in the calculation.
As a result, the decay width is
\begin{eqnarray}\label{eq:Gamma}
\Gamma=\frac{\pi|\boldsymbol{P}|}{4M_A^2}\sum_{LS}|\mathcal{M}^{LS}(\boldsymbol{P})|^2
\end{eqnarray}

As one parameter of the model, we adopt $\gamma=8.77$ from Ref.~\cite{Close:2005se},
which is almost the same as the value in related works \cite{Li:2008et,Ye:2012gu,He:2013ttg}.
The other is the SHO wavefunction scale parameter $\beta$, which represents the effective radius of
the particle ($R\sim 1/\beta$). Although conventionally $\beta$=0.4 GeV for all decay modes, we apply
varying scale parameters for states with different quantum numbers ($1S, 1P, 1D, 2S, 3S$) as in Ref.~\cite{Close:2005se},
where the $\beta$ in the SHO of produced particles, mainly the lower excitation state (the principal quantum number $n \leq 3$),
is determined by matching it to reproduce the root mean square radius predicted by a nonrelativistic quark model with
Coulomb + linear and Gaussian-smeared hyperfine interactions. Note that in Ref.~\cite{Close:2005se},
the atomic notation of $n\,L$ is used, and we adapt it as $n+1\, L$ to be consistent with our standard convention for mesons.
On the other hand, we try to obtain the parameter for the wavefunction of the highly excited mother meson by fitting the decay
width of 2$^3S_1$ mesons with the latest experimental data (Table~\ref{tab:table1}). The value we obtain is
$R$=4.34 $\text{GeV}^{-1}$ or $\beta$=0.23 GeV, which is actually very close to the value in Ref.~\cite{Close:2005se} and
thus might justify our choice.
\begin{table}[h]
\caption{\label{tab:table1}
Determine the parameter $\beta$ of the SHO wavefunction for mother mesons from the widths of 2$^3S_1$
mesons by minimizing $\chi^2$ (the square difference between the numerical result of the model and
the data from the latest experiments). The final result is $\beta$=0.23 GeV. Since the data of
different experiments differ greatly, with various peaks, we cite only the latest results.
We do not use $K^{*}(1410)$ for the fitting due to some problems in assigning it as the 2$^3S_1$ state, as discussed later.
}
\begin{ruledtabular}
\begin{tabular}{ccc}
Particle &Fit result (MeV) & Experiment (MeV)\\
\colrule
$\rho$(1450)&271&280$\pm$20 (SND 18)\cite{Achasov:2017kqm}\\
$\omega$(1420)&135&104$\pm$35$\pm$10 (CMD3 17)\cite{CMD-3:2017tgb}\\
$\phi$(1680)&77&$103^{+26}_{-24}$ (SND 19)\cite{Achasov:2019duv}\\
\end{tabular}
\end{ruledtabular}
\end{table}

\section{{\label{sec:level3}The $\rho$ and $\omega$ mesons}}
\begin{figure}[htbp]
\includegraphics[width=0.45\textwidth]{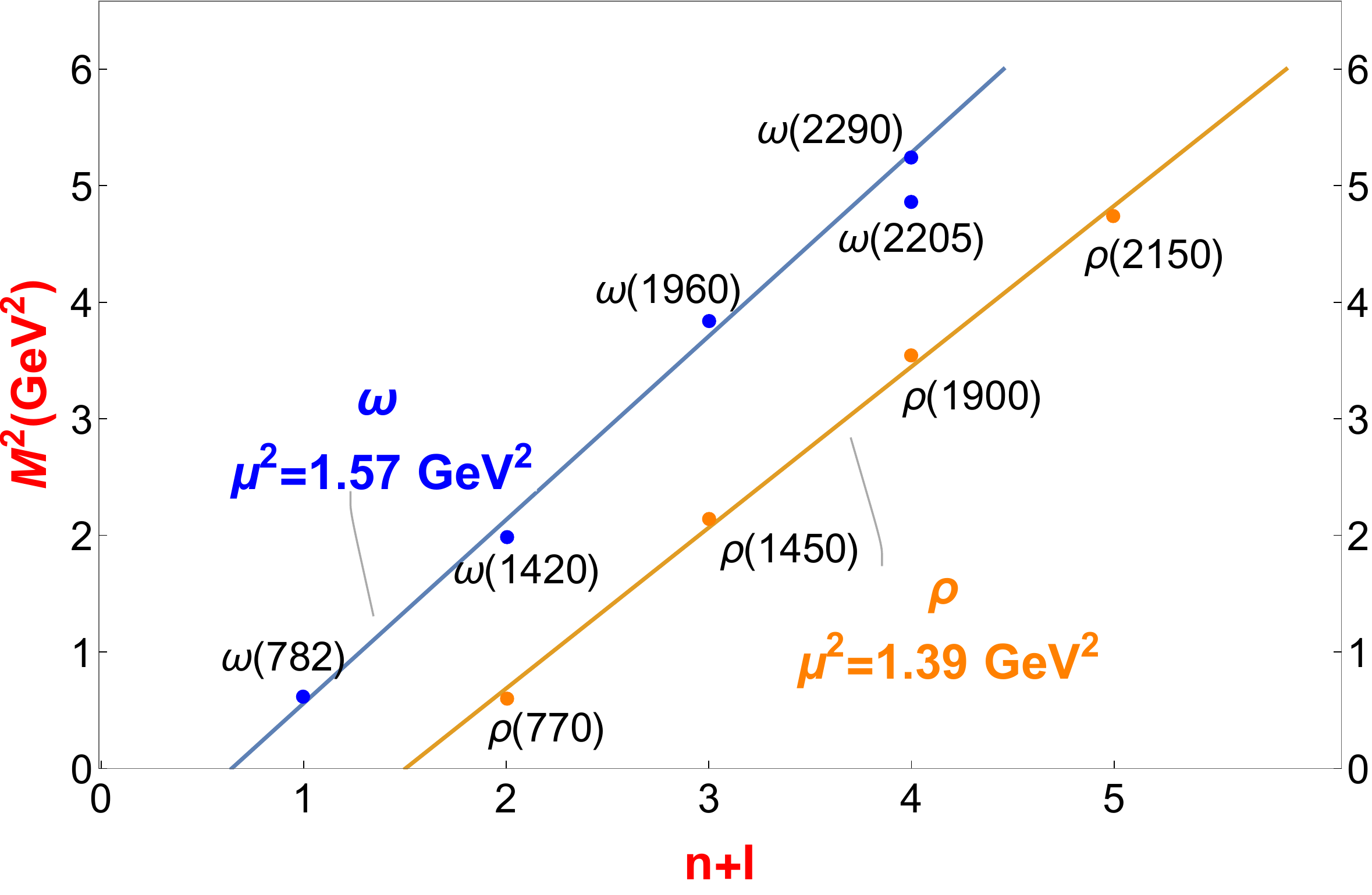}
\caption{\label{fig:epsart1} $\mu^2$ is the slope of the line. To present the two families clearly,
the horizontal axis is $n+I$ instead of $n$. The measured masses of $\rho(2150)$ are inconsistent; hence,
we simply adopt the average of the central value 2176 MeV from 4 recent experiments to perform the following calculations \cite{Zyla:2020zbs}}.
\end{figure}

Use of the Regge trajectory is a simple and effective approach to study the mass spectrum of mesons
with sufficient accuracy \cite{Anisovich:2000kxa,Masjuan:2012gc}. By applying this approach, our calculation
is much simplified, compared to some quark models with solving Sch\"ordinger-like equation.
As a result, the higher excited states are predicted based on the well-established lower states from experiment.
In this sense, our consideration could be a correct but also economic one.
According to the Regge trajectories and predictions of other works on the quark model \cite{Godfrey:1985xj,Ebert:2009ub},
there are some possible candidates for the $\rho$ and $\omega$ meson families, with quantum numbers of 3$^3S_1$ and 4$^3S_1$.
We list them in Table~\ref{tab:table2}. The corresponding Regge trajectories are plotted in Fig.~\ref{fig:epsart1}.
\begin{table}[htbp]
\caption{\label{tab:table2}
Observed $\rho$ and $\omega$ candidates, judged by their masses.
}
\begin{ruledtabular}
\begin{tabular}{ccc}
&3$^3S_1$ &4$^3S_1$\\
\colrule
$I=1$ &$\rho(1900)$ &$\rho(2150)$\\
$I=0$ &$\omega(1960)$ &$\omega(2205)$, $\omega(2290)$\\
\end{tabular}
\end{ruledtabular}
\end{table}

\subsection{$\rho$ meson}

\subsubsection{$\rho(1900)$}
The state $\rho(1900)$ is a natural candidate for the second radial excitation of the ground state $\rho(770)$,
determined based on its mass. We calculated the strong two-body decay widths shown in Table~\ref{tab:table3}.
The theoretical total width is approximately one $\sigma$ around the BaBar experimental value \cite{Aubert:2006jq},
which is regarded as a supporting indication of the 3$^3S_1$ assignment. The main decay modes predicted are $\pi\pi$,
$a_1\pi$, and $\omega \pi$, which suggest a noticeable 4$\pi$ mode that has not been previously seen.

However, we have to consider the inconsistency in the measurement of the width of $\rho(1900)$. Actually,
two width values are observed in the experiments: one around 150 MeV and another below 50 MeV. This result
may be caused by the irregular behavior of the cross-sections near the $N\bar{N}$ threshold, as suggested
 by the PDG \cite{Zyla:2020zbs}. In some recent theoretical works, dips around 1900 MeV can be explained by
 the influence of the $N\bar{N}$ channel, the threshold of which is 1.88 GeV
 \cite{Haidenbauer:2015yka,Kang:2013uia,Haidenbauer:2014kja,Dai:2017ont,Milstein:2018orb}. If this is the case,
 then this effect is beyond the quark pair creation model, and it is reasonable to compare our result only with the relatively broader width.

\subsubsection{$\rho(2150)$}\label{sec:rho2150}
Although the relativized quark model of Godfrey and Isgur predicts the $2 ^3D_1$ states at 2150 MeV \cite{Godfrey:1985xj},
the Regge analysis of both our model and those of others suggests that $\rho(2150)$ is more preferable as a candidate of 4$^3S_1 $
\cite{Anisovich:2000kxa,Masjuan:2012gc}. Our calculation of the decay width with the $^3P_0$ model, the results of which are given
in Table~\ref{tab:table3}, could support this argument. For the experimental data, there is a large difference between the widths
measured before the 2010s, which are above 300 MeV, and the latest results of 70-190 MeV; more details can be found in the publication
by the PDG. The result of our theoretical prediction is 116 MeV, matching most of the latest observations within one $\sigma$ \cite{Zyla:2020zbs}.
Moreover, the numerical result of the partial widths can partly explain the observed modes, which are $\pi^+\pi^-$, $K^+K^-$, $6\pi$,
$\eta'\pi^+\pi^-$, $f_{1}(1285)\pi^+\pi^-$, $\omega\pi^0$ and $\omega\pi^0\eta$.

We have noticed that in the past two years, there have been more precise measurements of resonances around 2.1 GeV by BaBar and
BES$\text{\Rmnum{3}}$. The first structure was observed in the process of $e^+e^-\to K^+K^-$ by BES$\text{\Rmnum{3}}$ in 2019,
which has a mass of 2239.2$\pm7.1\pm11.3$ MeV and a width of 139.8$\pm12.3\pm20.6$ MeV. Possible candidates for it are $\rho(2150)$ and
$\phi(2170)$ \cite{Ablikim:2018iyx}. Later, BaBar confirmed its existence, combining more data from other channels, and described it by a
model of $\rho(2230)$ with mass and width $M$=2232$\pm8\pm9$ MeV and $\Gamma$=$133\pm14\pm4$ MeV, respectively \cite{BABAR:2019oes}.
In the following year, BES$\text{\Rmnum{3}}$ observed another structure, called $Y(2040)$, in the $\omega\pi^0$ cross section, with a mass
of 2034$\pm13\pm9$ MeV and a width of 234$\pm30\pm25$ MeV \cite{Ablikim:2020das}. From the latest data for $e^+e^-\to \eta'\pi\pi$
of BES$\text{\Rmnum{3}}$ in December 2020, a resonance with a mass and width of $M$=2108$\pm46\pm25$ MeV and $\Gamma$=138$\pm36\pm30$ MeV
was shown to match $Y(2040)$ \cite{Ablikim:2020wyk}.

In fact, it is very difficult to conclude whether there exist two peaks around 2.1 GeV or just one state from the data we have.
If they are confirmed to be one state, our calculation of the total width favors the 4$^3S_1$ assignment. However, the inconsistency
in their masses may be explained by the existence of more than one structure, one with a mass of approximately 2.1 GeV for $\rho(4^3S_1)$
and another with a mass of approximately 2.2 GeV for $\omega(4^3S_1)$, but with similar widths (see also the subsection on $\omega(4^3S_1)$ below).
We also notice that the $\rho$ and $\omega$ mesons have different isospins, and thus, a combined analysis of the final states $\omega\pi^0$ and
$\omega\eta$ is helpful to discriminate the $\rho$ and $\omega$ families.

\begin{table}[htbp]
\caption{\label{tab:table3}
Partial widths of $\rho(1900)$ as 3$^3S_1$ and $\rho(2150)$ as 4$^3S_1$. The data of the total width are from Refs.~\cite{Aubert:2006jq,BABAR:2019oes}.
There is no corresponding value for the mode prohibited by the phase space. Numbers with two decimal points are displayed as 0.
}
\begin{ruledtabular}
\begin{tabular}{ccc}
&$\rho(1900)$&$\rho(2150)$\\
 \cline{2-3}
Mode&Width (MeV)&Width (MeV)\\
\colrule
$\pi\pi$&70&33 \\
$\pi(1800)\pi$& &25\\
$a_1(1260)\pi$&49&14\\
$\omega(1420)\pi$&22&7\\
$b_1(1235)\rho$& &8\\
$h_1(1170)\pi$&10&3\\
$\omega\pi(1300)$& &5\\
$\rho\rho$&17&2\\
$f_2(1270)\rho$& &4\\
$a_2(1320)\omega$& &3\\
$a_1(1260)\omega$& &2\\
$KK$&5&2\\
$\eta(1295)\rho$& &1\\
$f_1(1285)\rho$& &1\\
$KK_1(1270)$&2&1\\
$\pi_2(1670)\pi$&4&1\\
$K^*(1410)K$& &1\\
$b_1(1235)\eta$&1&0.5\\
$KK_1(1400)$&0.2&0.4\\
$\eta'\rho$&0.7&0.3\\
$\eta\rho$&0.5&0.3\\
$\pi(1300)\pi$&0&0.3\\
$\omega(1650)\pi$&0.9&0.2\\
$\omega_3(1670)\pi$&0&0.2\\
$\omega\pi$&1&0.2\\
$a_2(1320)\pi$&0.5&0.2\\
$\eta\rho(1450)$& &0.1\\
$KK^*$&0&0.1\\
$K^*K^*$&0&0\\
$KK^*_2(1430)$& &0\\ \colrule
Total&184&116
\\\colrule
Experiment &160$\pm$20(BABAR 06) &127$\pm$14$\pm$4
\end{tabular}
\end{ruledtabular}
\end{table}

\subsection{$\omega $ meson}
\subsubsection{$\omega(1960)$}
Only one state is observed in the possible mass region of $\omega(3^3S_1)$ calculated by the Regge trajectories.
$\omega(1960)$ has been discovered only once by the SPEC and is therefore listed in ``Further States''.
The total decay width produced by the $^3P_0$ model is 217 MeV, and the dominant modes are $\rho(1450)\pi$, $b_1(1235)\pi$ and $\rho\pi$,
which contribute more than 90\% (Table~\ref{tab:table4}). The theoretical value is close to the experimental width of 195$\pm$60 MeV.
In addition, the resonance is clearly visible in the $b_1\pi$ channel in the experiment \cite{Anisovich:2011sva},
as expected from our calculation. With the fact that it couples to $^3D_1$, we cannot preclude the possibility of 2$^3D_1$ being assigned,
as has been visualized in an explicit calculation \cite{Wang:2019jch}. However, the $\omega\eta$ data do produce the best determination of
the $1^{--}$ resonance at 1960 MeV \cite{Anisovich:2011sva}, challenging our tiny branching fraction of the $\omega\eta$ model.
We look forward to new experiments capable of separating more $1^{--}$ states out around 2 GeV. The partial width information is particularly valuable.
\begin{table}[h]
\caption{\label{tab:table4}
Partial widths of $\omega(1960)$ as 3$^3S_1$. The experimental data are from Ref.~\cite{Anisovich:2011sva}.
Numbers with two decimal points are displayed as 0.}
\begin{ruledtabular}
\begin{tabular}{cc|cc}
Mode &Width (MeV) &Mode &Width (MeV)\\\colrule
$\rho(1450)\pi$&85&$K^*(1410)K$&3\\
$b_1(1235)\pi$&83&$K_1(1400)K$&2\\
$\rho\pi$&22&$\omega\eta'$&1\\
$KK$&7&$\rho_3(1690)\pi$&0.6\\
$h_1(1170)\eta$&6&$K^*K$&0.1\\
$\rho(1700)\pi$&6&$K^*K^*$&0.1\\
$K_1(1270)K$&3&$\omega\eta$&0\\
\colrule
Total&\multicolumn{3}{c}{217}
\\\colrule
Experiment & \multicolumn{3}{c}{195$\pm$60}
\end{tabular}
\end{ruledtabular}
\end{table}

\subsubsection{$\omega(4^3S_1)$}
According to the Regge analysis, the mass of $\omega(4^3S_1)$ is approximately 2.2-2.3 GeV. Two candidates ($\omega(2205)$ and $\omega(2290)$)
are listed in ``Further States'' \cite{Anisovich:2011sva,Bugg:2004rj}. We also calculate an unobserved state $\omega(2319)$, which is predicted
directly by the Regge trajectory of the first three states of $\omega(n^3S_1)$, to provide more information. The total decay widths,
all approximately 140 MeV, vary slightly with the mass (Table~\ref{tab:table5}). The main decay modes are $b_1\pi$, $\rho\pi$ and $a_2\pi$,
emphasizing the importance of analyzing the data of the $\omega\pi\pi$ and $\rho\pi\pi$ channels. In addition, the $KK$ mode might be measurable.
As mentioned in Sec.~\ref{sec:rho2150}, we noted that the $^3P_0$ model produced a similar width for $\rho(4^3S_1)$ and $\omega(4^3S_1)$,
116 MeV and 140 MeV respectively, which both agree with the results of recent experiments to some degree of
uncertainty \cite{Ablikim:2018iyx,BABAR:2019oes,Ablikim:2020das,Ablikim:2020wyk}.

There is a large difference between our result and the two sets of experimental data. This difference might be an indication of exotic states,
which means that more effects beyond the $q\bar{q}$ meson model need to be considered. Alternatively, the problem might be the result of the
blurred combination of two states in the mass region of 2110-2230 MeV, as already indicated in Ref.~\cite{Anisovich:2011sva}. Moreover,
this resonance should have a large $^3D_1$ amplitude. A partial-wave analysis is indispensable to pin it down. We also notice that the
polarized target experiment provides more accurate polarization information for the daughter particle, which is very advantageous as it
provides access to the interferences between partial waves; thus, the partial wave information and the resonance parameters could be more
precisely measured \cite{Anisovich:2002su}.

Surely, another way to improve the situation is to perform more precise measurements of the partial widths to distinguish the states.
In addition, we have to apply more accurate models beyond $^3P_0$ to make more reliable predictions. An explicit example is to utilize
the relativistic quark model based on the quasipotential approach \cite{Ebert:2014jxa}. In that calculation, the relativistic
structure of the decay matrix element, relativistic contributions and boosts of the
meson wave functions are comprehensively taken into account. Thus, it certainly extends beyond the simplified $^3P_0$ model.
For the isospin part, a direct and clear solution is to measure various final states with definite isospins.

Interestingly, we add a few comments here. For a specific decay channel of the mother particles with the same quantum number,
the inputs of $\beta$ (of course also including the universal $\gamma$) are the same, and one may then naturally expect
that the mother particle with a heavier mass will have the larger decay width due to larger phase space available. However, it is not always so.
An explicit example is the decays $\omega(2205),\omega(2290),\omega(2319)\to \rho(1450)\pi$ in Table \ref{tab:table5}.
In fact, we should also consider the dynamical matrix element. The function of $I(\boldsymbol{P},m_1,m_2,m_3)$ in Eq.~\eqref{eq:I} also depends
on $\boldsymbol{P}$. As a result, the decay width $\Gamma$ (see Eq.~\ref{eq:Gamma}) is not a monotonic function of $|\boldsymbol{P}|$.
As an explicit example, one may refer to Fig.~4 in Ref.~\cite{Li:2020xzs}, where the width for the channel $K K_1(1400)$ drops dramatically
with increasing mass.
The nodes of wave function influence the predicted width. Additionally, we also notice the $^3P_0$ model
and a more realistic model may lead to very different predictions for the width of a specific channel,
e.g. Table VI in Ref.~\cite{Xue:2018jvi}, but the prediction on the total width is generally reasonable and acceptable.

\begin{table}[htbp]
\caption{\label{tab:table5}
Partial widths of the candidates of $\omega$ 4$^3S_1$. The experimental data are from Refs.~\cite{Anisovich:2011sva,Bugg:2004rj}.
Numbers with two decimal points are displayed as 0.
}
\begin{ruledtabular}
\begin{tabular}{cccc}
&$\omega$(2205)&$\omega$(2290)&$\omega$(2319)\\
\cline{2-4}
Mode &Width (MeV) &Width (MeV) &Width (MeV)\\\colrule
$b_1(1235)\pi$&24&35&41\\
$\rho(1450)\pi$&30&26&21\\
$a_2(1320)\rho$&40&23&10\\
$\rho\pi$&4&16&22\\
$\rho\pi(1300)$&5&6&15\\
$KK$&2&3&4\\
$K_1(1400)K$&1&3&2\\
$f_1(1285)\omega$&3&1&2\\
$f_2(1270)\omega$&10&1&0.3\\
$\rho(1700)\pi$&1&4&5\\
$h_1(1170)\eta$&4&6&7\\
$a_0(1450)\rho$&0&4&0.3\\
$a_1(1260)\rho$&12&4&2\\
$\omega(1420)\eta$&1&4&4\\
$K^*(1410)K$&0.5&1&1\\
$K_1(1270)K$&0.6&0.9&1\\
$f_0(1370)\omega$&0.5&0.8&1\\
$\omega\eta(1295)$&0&0.7&0.1\\
$\omega\eta'$&0.4&0.4&0.3\\
$\rho_3(1690)\pi$&0.8&0.3&0.1\\
$\omega\eta$&0&0.1&0.3\\
$K^*K$&0&0.1&0.1\\
$K^*K^*$&0&0.1&0.1\\
$K^*(1680)K$&0&0&0\\\colrule
Total&141&138&142
\\\colrule
Experiment &350$\pm$90 (SPEC 02)&275$\pm$35 &
\end{tabular}
\end{ruledtabular}
\end{table}

\section{{\label{sec:level4}The $\phi$ and $K$ mesons}}
\subsection{$\phi$ meson}

\begin{figure}[htbp]
\includegraphics[width=0.45\textwidth]{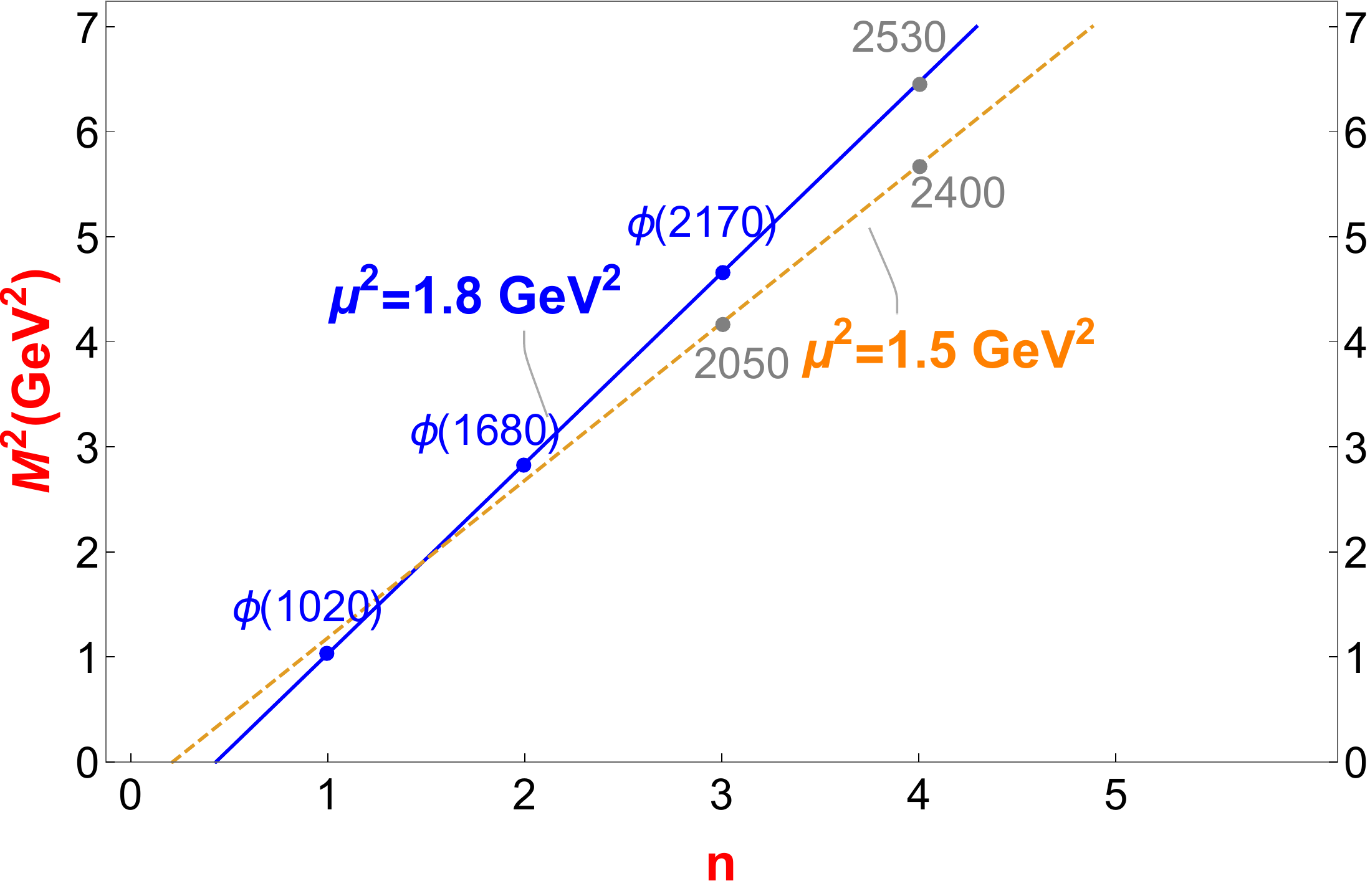}
\caption{\label{fig:epsart2} The straight line $\mu^2$=1.8 $\text{GeV}^2$ treats $\phi(2170)$ as $3^3S_1$,
and the dashed line adopts the average slope of $\rho$ and $\omega$ of approximately 1.5 $\text{GeV}^2$.
The blue dots are the observed states, and the gray dots are the predicted states.}
\end{figure}

Unlike $\rho$ and $\omega$, only one state $\phi(2170)$ above the first radial excitation $\phi(1680)$ has been found experimentally.
If it is the 3$^3S_1$ state of the $\phi$ meson, the slope of the Regge trajectory will be $\mu^2$=1.8 $\text{GeV}^2$, which is greater
than that of the $\rho$ and $\omega$ families and actually is higher than that of all nonstrange light $q\bar{q}$ mesons \cite{Masjuan:2012gc}.
Therefore, we draw two trajectories in Fig.~\ref{fig:epsart2}, one with a slope $\mu^2$=1.8 $\text{GeV}^2$ and another with the average
slope of $\rho$ and $\omega$, i.e., approximately 1.5 $\text{GeV}^2$, to predict the mass of $\phi(3^3S_1)$ and $\phi(4^3S_1)$.

First, we calculate the $\phi(2170)$ decay width with the $^3P_0$ model, along with that of the predicted $\phi(2050)$
by the less steep line (Table~\ref{tab:table6}). As we can see, the theoretical value is much higher than the experimental data.
This discrepancy has previously been observed by others using the $^3P_0$ model, which might imply the incorrectness of assigning
$\phi(2170)$ as a pure $q\bar{q}$ $3^3S_1$ state \cite{Barnes:2002mu,Pang:2019ttv,Li:2020xzs}. In addition, we can compare the partial
widths with the latest data of BES$\text{\Rmnum{3}}$ \cite{Ablikim:2020pgw}. The result of our calculation is that
$\Gamma(K_1(1400)K)\simeq\Gamma(K_1(1270)K)>\Gamma(K^*K^*)>\Gamma(K(1460)K)$, which does not agree with the experimental
observation of $\Gamma(K_1(1400)K)$ and $\Gamma(K(1460)K)$ with higher significance but limited significance in $K^*K^*$ and $K_1(1270)K$.
In conclusion, $\phi(2170)$ cannot be understood as a pure $3^3S_1$ state, and more interactions beyond the $q\bar{q}$ meson model are likely needed.

The corresponding $4^3S_1$ states predicted by the two trajectories are 2.4 GeV and 2.53 GeV. The partial widths are listed in Table~\ref{tab:table7}.
The total width of $\phi(2530)$ is too large to be reliable, which might indicate that $\phi(2400)$ with $\Gamma\simeq 240$ MeV is a better prediction
for $\phi(4^3S_1)$. We expect that the dominant modes are $KK_1(1270)$, $KK$, $KK_1(1400)$ and $K^*K$.

\begin{table}[htbp]
\caption{\label{tab:table6}
Partial widths of the observed $\phi(2170)$ and predicted $\phi(2050)$ as 3$^3S_1$.
The experimental mass and width data of $\phi(2170)$ are from the PDG average \cite{Zyla:2020zbs}.
}
\begin{ruledtabular}
\begin{tabular}{ccc}
&$\phi(2170)$&$\phi(2050)$\\
\cline{2-3}
Mode&Width (MeV)&Width (MeV)\\\colrule
$K_1(1400)K$&120&64\\
$K_1(1270)K$&111&71\\
$KK$&77&51\\
$K^*K$&55&14\\
$K^*K^*$&20&25\\
$h_1(1415)\eta$&15&8\\
$\phi\eta$&4&0.4\\
$K_2^*(1430)K$&3&0\\
$K^*(1410)K$&3&37\\
$K(1460)K$&2&6\\
$\phi\eta'$&0.6&0.4\\
\colrule
Total&410&277\\
\colrule
Experiment & 125$\pm$65&
\end{tabular}
\end{ruledtabular}
\end{table}

\begin{table}[htbp]
\caption{\label{tab:table7}
Partial widths of the predicted $\phi(2400)$ and $\phi(2530)$ as 4$^3S_1$.
We do not include all possible two-body modes for $\phi(2530)$ because the
rest modes have a mass close to 2530 MeV and, more importantly, the total width of $\phi(2530)$ is too wide to be reliable.
}
\begin{ruledtabular}
\begin{tabular}{ccc}
&$\phi(2400)$&$\phi(2530)$\\
\cline{2-3}
Mode&Width (MeV)&Width (MeV)\\\colrule
$KK_1(1270)$&61&108\\
$KK$&49&79\\
$KK_1(1400)$&45&69\\
$K^*K$&32&98\\
$K^*(1410)K^*$&0.4&110\\
$K^* K_1(1270)$&13&34\\
$\phi\eta$&7&27\\
$\eta h_1(1415)$&12&23\\
$K_2^*(1430)K$&3&25\\
$K^*K_1(1400)$&0&16\\
$K^*(1680)K$&4&12\\
$K_2(1770)K$&4&19\\
$K_2(1820)K$&2&18\\
$K^*(1410)K$&1&12\\
$K^*K(1460)$&1&12\\
$K^*K_2^*(1430)$&0&11\\
$K^*K^*$&5&2\\
$K^*K^*_0(1430)$&0&6\\
$KK(1460)$&0.6&3\\
$\eta'h_1(1415)$&0&3\\
$\phi f_0(1370)$&0&2\\
$\phi f_1(1420)$&&1\\
$K_3^*(1780)K$&0&1\\
$\phi\eta'$&0.7&0.2\\
\colrule
Total&241&691\\
\end{tabular}
\end{ruledtabular}
\end{table}

\subsection{Kaon}
\subsubsection{K$(2^3S_1)$}
As mentioned above, we do not assume $K^*(1410)$ as the 2$^3S_1$ state to fit the parameters due to some disagreements
on the $1^{--}$ state assignment. Including the most recent experiment on LHCb in 2016 \cite{Aaij:2016iza}, two $1^{--}$ states
heavier than the ground state $K^*(892)$ have been discovered: $K^*(1410)$, with mass $M=1414\pm15$ MeV and width $\Gamma=232\pm21$ MeV,
and $K^*(1680)$, with mass $M=1718\pm18$ MeV and a wide, uncertain average width $\Gamma=322\pm110$ MeV.

It seems natural to assign the lighter $K^*(1410)$ as 2$^3S_1$ and the heavier $K^*(1680)$ as 1$^3D_1$. However, this assignment
would contradict the prediction of most theoretical works, which expect the 2$^3S_1$ resonance to be around 1.6 GeV. On the other hand,
the mass of 1$^3D_1$ is predicted to be approximately 1.7 GeV, and therefore, $K^*(1680)$ has been regarded as 1$^3D_1$ to
a large extent \cite{Godfrey:1985xj,Ebert:2009ub}. However, the ratio of the partial width $\Gamma(K^*(1680)\to K \pi)$
to $\Gamma(K^*(1680)\to K\eta)$ disfavors $K^*(1680)$ as 1$^3D_1$ assignment \cite{Ablikim:2020coo}.

\begin{figure}[h]
\includegraphics[width=0.45\textwidth]{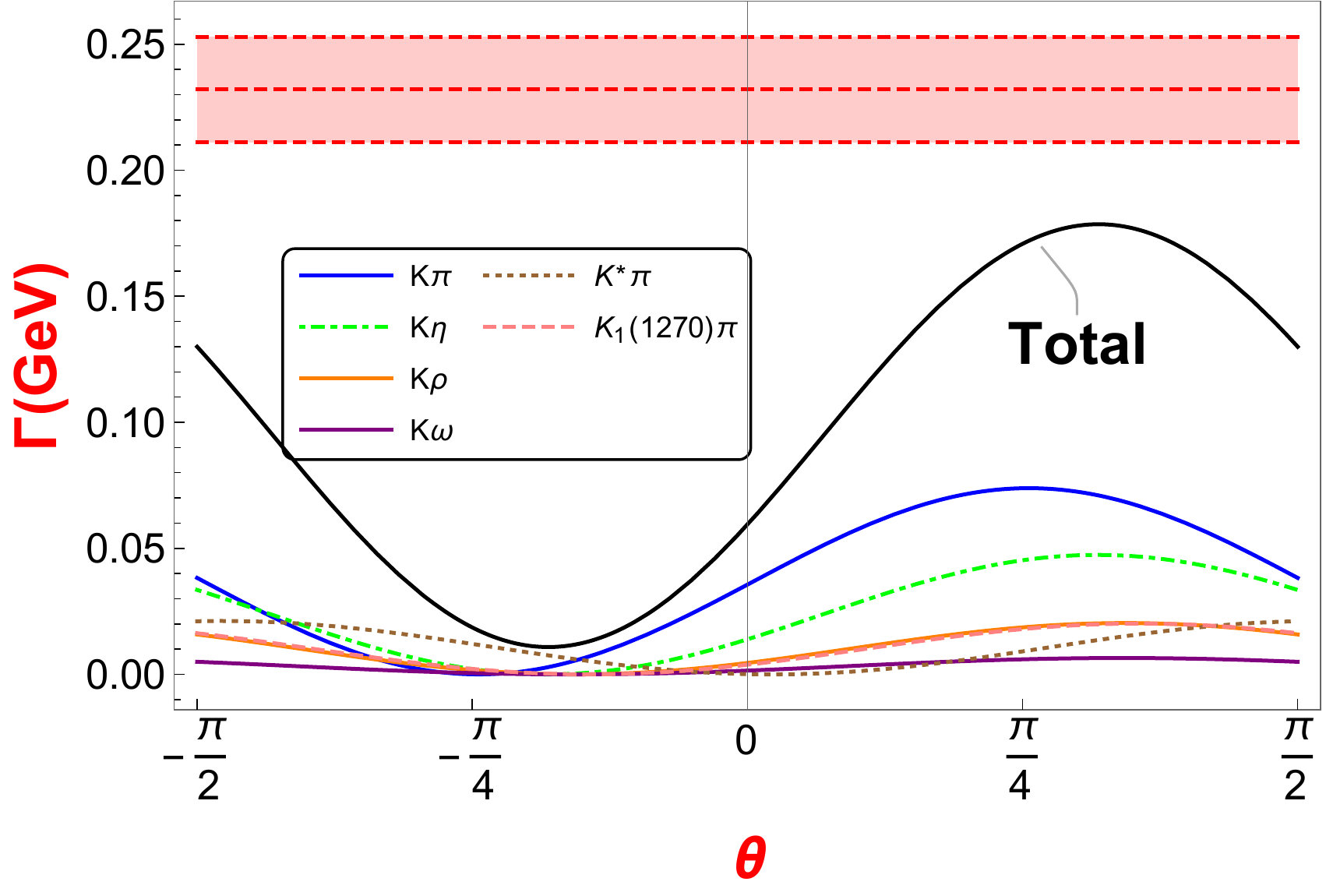}
\caption{\label{fig:epsart3}.The total width of $K^*(1410)$ varies with the mixing angle yet cannot reach the width of the experimental data.
The data from the PDG average with uncertainty $\pm\sigma$ is presented in red \cite{Zyla:2020zbs}.}
\end{figure}

The problem can also be analyzed with respect to the widths from the $^3P_0$ model. The first potential explanation for the discrepancy is that
$K^*(1410)$ is a mixture of 2$^3S_1$ and 1$^3D_1$:
\[\ket{K^*(1410)}=\cos\theta\ket{2^3S_1}+\sin\theta\ket{1^3D_1}\]
We calculate the dependence of the total and partial widths on the mixing angle $\theta$ in Fig.~\ref{fig:epsart3}.
Clearly, the theoretical value is lower than the experimental value by more than 2$\sigma$. The partial widths also
differ from our calculation. The branching fraction of the $K\pi$ mode is only $(6.6\pm1.0\pm0.8)\%$ \cite{Aston:1987ir},
 but it is one of the main modes predicted theoretically. Furthermore, $K^*\pi$ is observed with a branching fraction of
 more than 40\%, which contradicts our result, except around the minimum of the total width. Therefore, we do not assign
 $K^*(1410)$ as the $^3S_1$ state nor the mixture of 2$^3S_1$ and 1$^3D_1$. A similar doubt regarding the 2$^3S_1$ assignment
 has been expressed previously \cite{Barnes:2002mu}.

Inspired by some previous works, we examine the possibility that the resonance around 1.7 GeV consists of two peaks \cite{Burakovsky:1997ch},
similar to the argument made for $\rho(2150)$ and $\omega(4^3S_1)$. Despite the observations not showing clear inconsistency in the mass of
$K^*(1680)$, the hypothesis is still possible due to the limited data. Only three relatively accurate measurements are made: two from LASS
in 1987 and 1988 and one from LHCb in 2017. Their values are $1677\pm10\pm32$ MeV, $1735\pm10\pm20$ MeV and $1722\pm20^{+33}_{-109}$ MeV,
respectively \cite{Aston:1986jb,Aston:1987ir,Aaij:2016iza}. Therefore, we can argue that the lower part around 1680 MeV and the higher part
around 1730 MeV correspond to two $1^{--}$ states, $K^*(2^3S_1)$ and $K^*(1^3D_1)$, respectively, at least in theory.

We must emphasize that this idea is only a possibility without direct experimental evidence or indication but is derived from the problem
of assigning $K^*(1410)$ as 2$^3S_1$. Theoretical preference is the main motivation for the assignment. For example, compared to
$K^*(1410)$, $K^*(1680)$ is closer than the results predicted by the relativistic quark model,
which are 1580-1670 MeV \cite{Godfrey:1985xj,Ebert:2009ub}; if $K^*(1410)$ is 2$^3S_1$, it will be lightest
in the nonet with $\rho(1450)$ and $\omega(1420)$. In other words, the mass of $K^*(1410)$ is much lighter than the
theoretical predictions of the 2$^3S_1$ state of the kaon; however, $K^*(1680)$ is a little heavier.

\begin{figure}[htbp]
\includegraphics[width=0.45\textwidth]{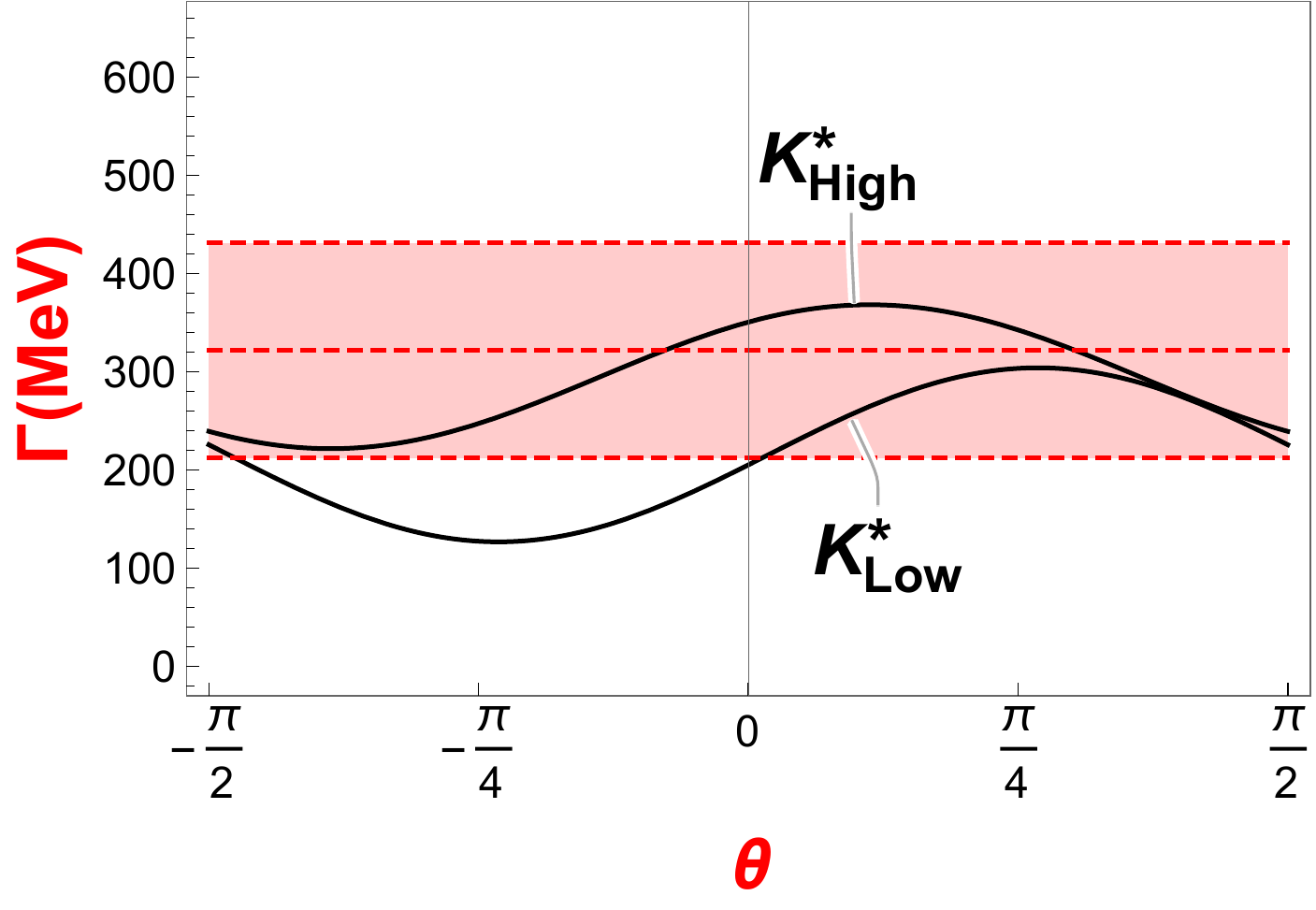}
\caption{\label{fig:epsart4}The total widths of $K^*_\text{High}$ ($M=1730$ MeV) and $K^*_\text{Low}$
($M=1680$ MeV) with variation in the mixing angle $\theta$. The PDG average
for $K^*(1680)$ with uncertainty $\pm\sigma$ is shown in red \cite{Zyla:2020zbs}.}
\end{figure}

\begin{figure}[htbp]
\includegraphics[width=0.45\textwidth]{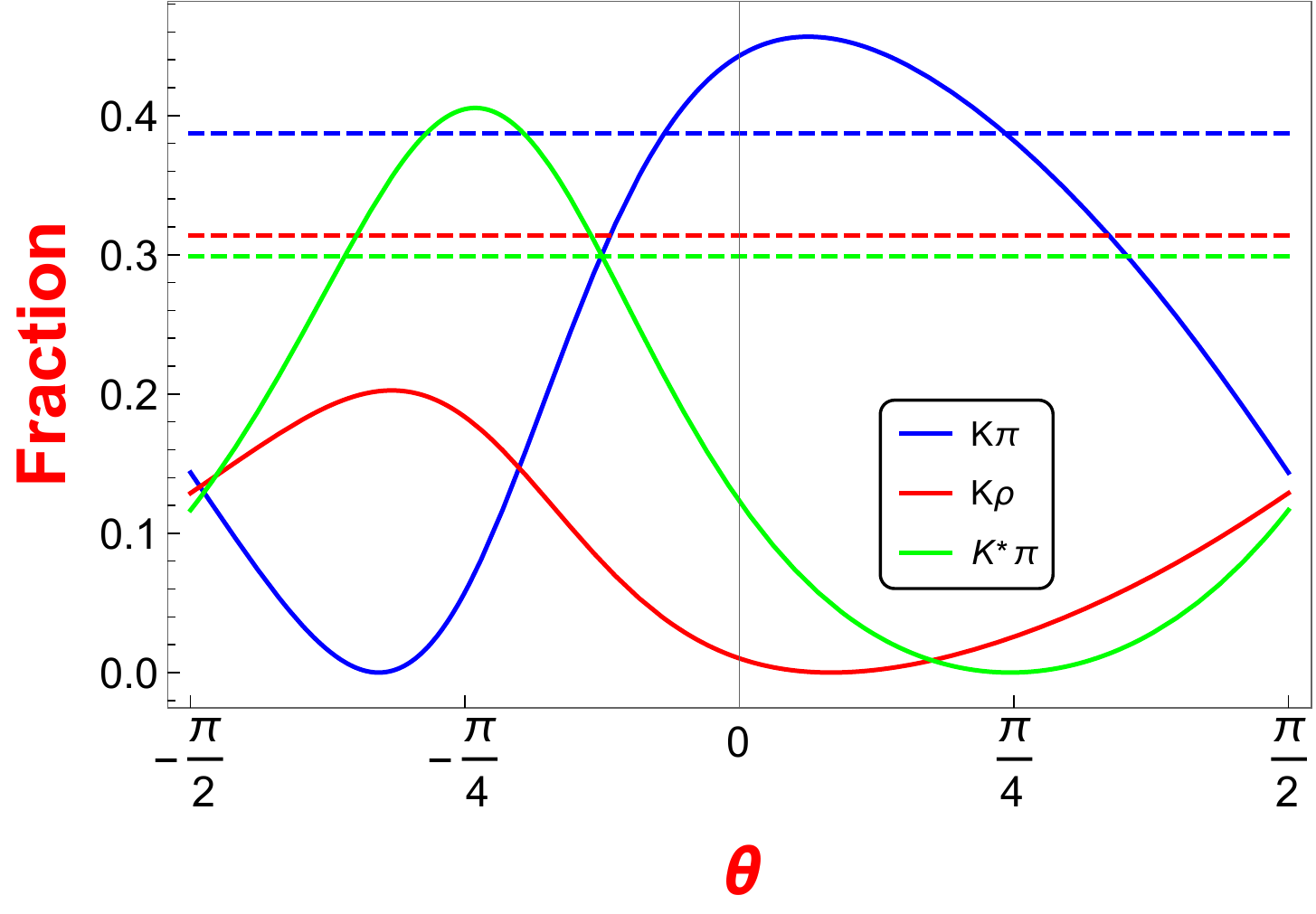}
\caption{\label{fig:epsart5}Three Branching fraction values are measured by LASS \cite{Aston:1986jb,Aston:1987ir}. The curves
represent the dependence of the branching fractions of modes $K\pi$, $K\rho$ and $K^*\pi$ on the mixing angle
for $K^*_\text{High}$. The dashed line with the same color denotes the corresponding experimental value from PDG.}
\end{figure}

\begin{figure}[htbp]
\includegraphics[width=0.45\textwidth]{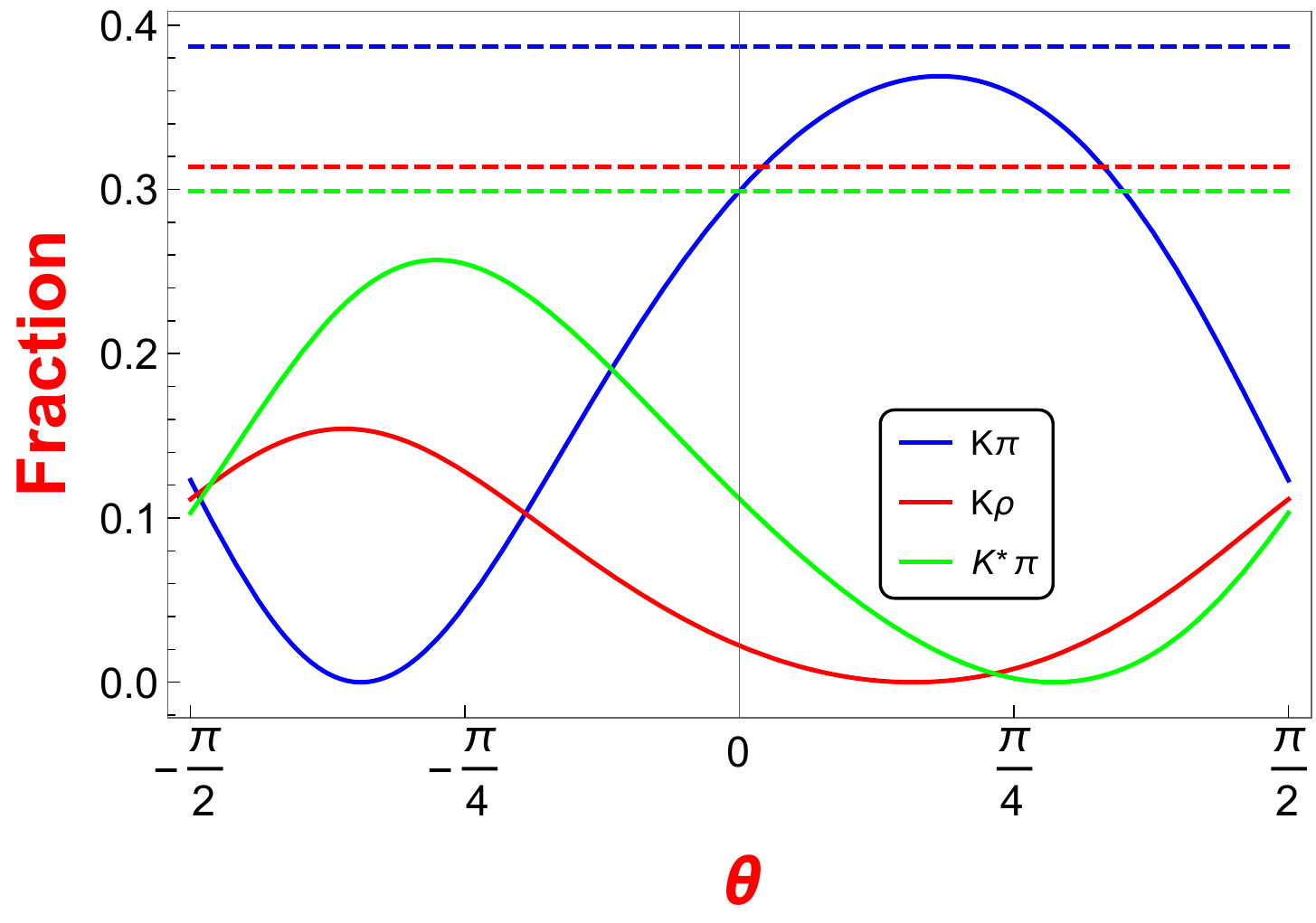}
\caption{\label{fig:epsart6} Comparison of the fractions for $K^*_\text{Low}$. The details are the same as in Fig.~\ref{fig:epsart5}.}
\end{figure}

To examine the idea more comprehensively, we calculate the decay width of $K^*(1680)$ in Table~\ref{tab:table8}
and make a comparison with experiments in Fig.~\ref{fig:epsart4}. In the calculation, the mass of the higher state is 1730 MeV,
and the mass of the lower state is 1680 MeV. To avoid the confusion between the name of the observed structure, $K^*(1680)$, and
our two hypothesized states, we call the hypothesized states $K^*_\text{High}$ and $K^*_\text{Low}$. We assume both of them to be
mixtures of 2$^3S_1$ and 1$^3D_1$, which might help us to compare the two assignments:
\[\ket{K^*_\text{High/Low}}=\cos\theta\ket{2^3S_1}+\sin\theta\ket{1^3D_1}.\]
Of course, the mixing angles $\theta$ for $K^*_\text{High}$ and $K^*_\text{High}$ are not necessarily the same.
The total width of $K^*_\text{High}$ can explain the data as $\theta$ varies in most of the region, but the best-fitted values of
the mixing angle are $-20^{\circ}$ and $50^{\circ}$. The total width of $K^*_\text{Low}$ remains consistent with the experiment
as $\theta$ varies from $0^{\circ}$ to $90^{\circ}$, and the calculated width is close to the central value of the data when the
mixing angle is $45^{\circ}$. Thus, we prefer that both $K^*_\text{High}$ and $K^*_\text{Low}$ are mixtures of $2^3S_1$ and $1^3D_1$.
In addition, we plot the three fractions of the observed modes in Fig.~\ref{fig:epsart5} and Fig.~\ref{fig:epsart6} with the
values from the experiments. In the two similar figures, the fraction of $K\rho$ is much lower than the experimental data,
which might reflect the fundamental limitation of the $^3P_0$ model for some specific channels. However, as the mixing angle
$\theta$ varies from $0^{\circ}$ to $90^{\circ}$, the relative branching fractions are similar to the experimental result that
$\Gamma(K\pi)>\Gamma(K\rho)\simeq \Gamma(K^*\pi)$. We notice that some uncertainties due to $\beta$ may be reduced in
predicting the relative branching fractions. From this perspective, we also favor the concept that $K^*_\text{High}$ and
$K^*_\text{Low}$ are mixtures of $2^3S_1$ and $1^3D_1$ with a positive mixing angle. Precise determination of the mixing
angle requires more accurate data.

In conclusion, if there are two structures around 1.7 GeV, assigning them as $K(2^3S_1)$ and $K(1^3D_1)$ is a more
favorable explanation than assigning $K^*(1410)$ as $2^3S_1$, according to their masses. The total decay widths and
relative values of the branching fraction from the $^3P_0$ model also support the idea that the two hypothesized states
are mixtures of $K(2^3S_1)$ and $K(1^3D_1)$.

\begin{table}[htbp]
\caption{\label{tab:table8} Partial widths of $K^*_\text{Low}$ ($M=1680$ MeV) and $K^*_\text{High}$
($M=1730$ MeV) with variation in the mixing angle (c means cosine and s means sine).
}
\begin{ruledtabular}
\begin{tabular}{ccc}
&$K^*_\text{Low}$&$K^*_\text{High}$\\
\cline{2-3}
Mode&Width (MeV)&Width (MeV)\\\colrule
$K\pi$&$91c^2+109cs+32s^2$   &$105c'^2+111c's'+29s'^2$\\
$K\eta$&$54c^2+96cs+43s^2$   &$66c'^2+104c's'+41s'^2$\\
$K\rho$&$2c^2-16cs+29s^2$   &$8c'^2-28c's'+26s'^2$\\
$K\omega$&$0.5c^2-5cs+10s^2$   &$2c'^2-9c's'+9s'^2$\\
$K^*\pi$&$25c^2-52cs+26s^2$   &$39c'^2-62c's'+35s'^2$\\
$K\eta'$&$0.3c^2+0.9cs+0.9s^2$   &$0.4c'^2+1c's'+1s'^2$\\
$K^*\eta$&$0.001c^2-0.1cs+0.9s^2$   &$0.05c'^2-0.4c's'+1s'^2$\\
$K\phi$&$0.2c^2-2cs+4s^2$   &$0.6c'^2-4c's'+6s'^2$\\
$Kh_1(1170)$&$10c^2+46cs+54s^2$  &$8c'^2+44c's'+62s'^2$\\
$K_1(1270)\pi$&$1c^2+9cs+15s^2$   &$0.3c'^2+3c's'+7s'^2$\\
$K_1(1400)\pi$&$7c^2-5cs+9s^2$   &$5c'^2-5c's+8s'^2$\\
$K^*\rho$&$12c^2-6cs+2s^2$   &$90c'^2-47c's'+18s'^2$\\
$K^*\omega$&$1c^2-0.5cs+0.2s^2$  &$26c'^2-14c's'+5s'^2$\\
\colrule
Total&$205c^2+176cs+226s^2$  &$351c'^2+95c's'+239s'^2$\\
\end{tabular}
\end{ruledtabular}
\end{table}

\begin{table}[htbp]
\caption{\label{tab:table9}
Partial widths of the predicted $K^*(1950)$ and $K^*(2230)$ as 3$^3S_1$.}
\begin{ruledtabular}
\begin{tabular}{ccc}
&$K^*(1950)$&$K^*(2230)$\\
\cline{2-3}
Mode&Width(MeV)&Width(MeV)\\\colrule
$K\pi$&40&99\\
$K\pi(1300)$&3&9\\
$K(1460)\pi$&0.2&9\\
$K\eta$&22&69\\
$K\eta'$&0.1&0.7\\
$K\eta(1295)$&0&0.3\\
$K\eta(1475)$&&0\\
$K(1460)\eta$&&0\\
$K\rho$&0.1&33\\
$K\rho(1450)$&&8\\
$K\omega$&0&10\\
$K\omega(1420)$&5&0.2\\
$K\phi$&0&5\\
$K\omega(1680)$&&0.4\\
$K^*\pi$&7&70\\
$K^*(1410)\pi$&1&23\\
$K^*\pi(1300)$&0&25\\
$K^*\eta$&0&0.4\\
$K^*\eta'$&1&0\\
$K^*(1410\eta$&&0.4\\
$K^*\rho$&0.7&0.2\\
$K^*\omega$&0.3&0.1\\
$K^*\phi$&0&0.2\\
$b_1(1235)K$&9&34\\
$h_1(1170)K$&13&49\\
$h_1(1415)K$&0&0.9\\
$K_1(1270)\pi$&27&70\\
$a_1(1245)K$&13&38\\
$f_1(1285)K$&0.6&2\\
$f_1(1420)K$&0&26\\
$K_1(1400)\pi$&30&64\\
$K_1(1400)\eta$&&44\\
$a_2(1320)K$&1&5\\
$f_2(1270)\eta$&0&2\\
$f_2(1525)\eta$&&0.4\\
$K_2^*(1430)\pi$&1&27\\
$K_2^*(1430)\eta$&&2\\
$K_2(1770)\pi$&0.4&13\\
$K^*(1680)\pi$&1&5\\
$K_2(1820)\pi$&&14\\
$K_3(1780)\pi$&0&4\\
\colrule
Total&176&760\\
\end{tabular}
\end{ruledtabular}
\end{table}

\subsubsection{K$(3^3S_1)$}
Similar to the prediction of the next unobserved $^3S_1$ state for the $\phi$ meson, we calculated the mass of $K(3^3S_1)$
from Regge trajectories with two slopes. If we use the average slope of $\rho$ and $\omega$, which is 1.5 $\text{GeV}^2$,
the next radial excitation is approximately 1.95 GeV. If we simply connect the $K^*$ and $K^*(1410)$ states, the next $^3S_1$
state will be 1.78 GeV, only slightly above $K^*(1680)$, and the width should not notably change. Alternatively, we can connect
$K^*$ and $K^*(1680)$ by postulating the latter as $2^3S_1$ to determine that $3^3S_1$ is approximately 2.23 GeV. We calculate
the total and partial decay widths of $K(3^3S_1)$ with a mass of 1.95 GeV or 2.23 GeV, the results of which are shown in
Table~\ref{tab:table9}. Their widths are 176 and 760 MeV, respectively. Since a very large width such as 700 MeV is not
very plausible for a resonant structure, the mass and width of $K(3^3S_1)$ are more likely $M\simeq2$ GeV and $\Gamma \simeq 200$ MeV,
respectively. Moreover, a previous work suggested $K(1830)$ as a good candidate for $3^1S_0$, with a mass around 1.87 GeV \cite{Barnes:2002mu}.
Therefore, 1950 MeV seems to be a more reasonable prediction, which is just a little heavier than the mass of $K(3^1S_0)$
due to the fine structure of the excited states. The main decay modes are $K\pi$, $K\eta$ and $K_1(1270)\pi$, indicating that
the $K\pi$, $K\eta$ and $K\pi\pi$ channels might be dominant in future experiments.

\section{{\label{sec:level5}Summary}}

We analyze the masses of excited light $^3S_1$ mesons with the Regge trajectory and calculate their two-body strong decay
widths with the $^3P_0$ model. By comparing the theoretical values with the latest experimental data, we provide several
suggestions regarding the assignment and some possible solutions to the current confusion.

First, the masses and widths of $\rho(1900)$ and $\omega(1960)$ match those of the experiments, and therefore, they are
respectively assigned to $\rho(3^3S_1)$ and $\omega(3^3S_1)$.

The inconsistent mass value for the measurement around the $\rho(2150)$ region might be explained by the two resonant structures
around 2.2 GeV. The $\rho(4^3S_1)$ and $\omega(4^3S_1)$ states are both predicted to be around 2.2 GeV with a similar total width of
approximately 110-140 MeV, which agree with the latest data from BaBar and BES$\text{\Rmnum{3}}$.

$\phi(2170)$ cannot be regarded as a pure $3^3S_1$ state due to the discrepancy in terms of both the total and partial widths
between the experiments and calculation. The $\phi(4^3S_1)$ state is predicted with a mass around 2.4 GeV and a width of 240 MeV.

Assigning $K^*(1410)$ as $2^3S_1$ contradicts the theoretical expectation of its mass and width. We try to explain the resonance
of $K^*(1680)$ by two peaks: $2^3S_1$ and $1^3D_1$. The total widths and branching fractions suggest that the two components of
$K^*(1680)$, i.e., $K^*_\text{High}$ and $K^*_\text{Low}$, could be mixtures of $2^3S_1$ and $1^3D_1$. This idea might be
theoretically plausible, but further evidence is needed. $K(3^3S_1)$ is predicted to have a mass of 1.95 GeV and a width of 180 MeV.

\begin{acknowledgments}
We thank Xian-Hui Zhong for the helpful discussions. We are grateful to Prof. V.~O.~Galkin for his careful reading and useful suggestions.
Various discussions with Wen-Biao Yan on the experimental measurements are also acknowledged.
This project is supported by the National Natural Science Foundation of China under Grant Nos. 11805012, 11635003, 11705056 and U1832173.
\end{acknowledgments}

\section*{References}
\bibliographystyle{apsrev}
\bibliography{BIB1}

\end{document}